\documentclass[10pt,conference]{IEEEtran}
\makeatletter
\usepackage{booktabs}
\usepackage{lipsum}
\newcounter{subsubsubsection}[subsubsection]
\renewcommand\thesubsubsubsection{\thesubsubsection.\@arabic\c@subsubsubsection}
\newcommand\subsubsubsection{\@startsection{subsubsubsection}{4}{\parindent}%
    {3.25ex plus 1ex minus .2ex}%
    {-1em}%
    {\normalfont\normalsize\bfseries}}
\newcommand*\l@subsubsubsection{\@dottedtocline{3}{10em}{4em}}
\newcommand*{\subsubsubsectionmark}[1]{}
\makeatother
\IEEEoverridecommandlockouts
\usepackage{graphicx}
\usepackage{cite}
\usepackage{listings}
\usepackage{tabularx}
\usepackage{booktabs}
\usepackage{amsmath,amssymb,amsfonts}
\usepackage{algorithmic}
\usepackage{graphicx}
\usepackage{textcomp}
\usepackage{hyperref}
\usepackage{xcolor}
\def\BibTeX{{\rm B\kern-.05em{\sc i\kern-.025em b}\kern-.08em
    T\kern-.1667em\lower.7ex\hbox{E}\kern-.125emX}}
\usepackage{graphicx}
\usepackage{array}
\usepackage{adjustbox}
\definecolor{darkgreen}{RGB}{0,100,0}
\usepackage{svg}

\begin{document}

\title{Imbalanced malware classification: an approach based on dynamic classifier selection\\
\thanks{This work was partially funded by Brazilian agencies: FACEPE and CNPq.}
}

\author{
\IEEEauthorblockN{José Vinicius S. Souza, Camila Barbosa Vieira, George D. C. Cavalcanti}
\IEEEauthorblockA{\textit{Centro de Informática} \\
\textit{Universidade Federal de Pernambuco}\\
Recife-PE, Brazil \\
\{jvss2, cbv2, gdcc\}@cin.ufpe.br}
\and
\IEEEauthorblockN{Rafael M. O. Cruz}
\IEEEauthorblockA{\textit{École de technologie supérieure} \\
\textit{Université du Québec}\\
Montreal, Canada \\
rafael.menelau-cruz@etsmtl.ca}
}

\maketitle

\begin{abstract}

In recent years, the rise of cyber threats has emphasized the need for robust malware detection systems, especially on mobile devices. Malware, which targets vulnerabilities in devices and user data, represents a substantial security risk. A significant challenge in malware detection is the imbalance in datasets, where most applications are benign, with only a small fraction posing a threat. This study addresses the often-overlooked issue of class imbalance in malware detection by evaluating various machine learning strategies for detecting malware in Android applications. We assess monolithic classifiers and ensemble methods, focusing on dynamic selection algorithms, which have shown superior performance compared to traditional approaches. In contrast to balancing strategies performed on the whole dataset, we propose a balancing procedure that works individually for each classifier in the pool. Our empirical analysis demonstrates that the KNOP algorithm obtained the best results using a pool of Random Forest. Additionally, an instance hardness assessment revealed that balancing reduces the difficulty of the minority class and enhances the detection of the minority class (malware). The code used for the experiments is available at \url{https://github.com/jvss2/Machine-Learning-Empirical-Evaluation}.
\end{abstract}


\section{Introduction}
In recent years, information security has become a key concern, with cybersecurity ranked as one of the top risks in both the short-term (two years) and long-term (ten years) \cite{GlobalRiskReport}. The rising frequency and sophistication of cyberattacks and data breaches heighten the threat to global digital security, particularly in regions with expanding internet access and weaker cybersecurity defenses \cite{GlobalRiskReport}. This gap between evolving threats and vulnerable infrastructures underscores the urgent need for more effective detection and prevention methods \cite{Aslan}.

%In recent years, information security has become a central concern across various sectors, with cybersecurity identified as one of the top risks in both the short-term (two years) and long-term (ten years), according to \cite{GlobalRiskReport}. The increasing frequency and sophistication of cyberattacks and data breaches intensify the danger to global digital security, especially in regions with expanding internet access where digital infrastructures are often less secure and lack advanced cybersecurity defenses \cite{GlobalRiskReport}. This disparity between evolving attack techniques and vulnerable infrastructures highlights the urgent need for more effective detection and prevention methods \cite{Aslan}.

A major digital security threat is malware, which refers to any software designed to harm a device, server, or network~\cite{MalwareDefinition}. On smartphones, malware can steal personal data, track locations, display ads, or even control the device remotely. Machine learning techniques have shown promise in malware analysis, offering efficient solutions for identifying patterns and anomalies in large datasets \cite{MLSurveyDetection}.
%One significant threat to digital security is malware. Malware, or malicious software, refers to any program or code developed with the intent to cause harm to a device, server, or network \cite{MalwareDefinition}. In the context of smartphones, malware can steal personal information, track user locations, display intrusive ads, or even remotely control the device. To combat these growing threats, machine learning techniques have shown promise in malware analysis, offering solutions capable of automatically and efficiently identifying patterns and anomalies in large datasets \cite{MLSurveyDetection}.

Malware analysis can employ a static approach, which examines the source or binary code without execution, searching for known signatures, code patterns, and other indicators of malicious intent \cite{MalwareAnalysisAproach}. While effective, this method struggles to identify new and unknown malware. Machine learning addresses this limitation by inferring new detection patterns; however, class imbalance, inherent to data collection, becomes a critical challenge. Rare instances are often misclassified due to imprecise detection rules, requiring careful attention to modeling and analysis \cite{ImbalancedReview}\cite{Bifet}.

This study presents an empirical analysis of machine learning algorithms for Android malware detection, comparing monolithic learning algorithms, static ensemble algorithms, and dynamic selection (DS) algorithms. DS algorithms dynamically select the most competent classifiers for each query instance \cite{DESRecentAdvances}, outperforming static combinations and monolithic classifiers, while effectively addressing imbalanced learning scenarios by adaptively focusing on the most relevant classifiers for rare instances, thus enhancing overall classification performance and robustness against minority class misclassification \cite{Cruz}.

The machine learning methods are evaluated on the Drebin dataset~\cite{Drebin}, a widely used benchmark for Android malware detection, which includes security-related variables such as access logs, network activity records, and intrusion indicators, enabling comprehensive analysis of malware behavior. To enhance performance, we propose a Bootstrap-Based Balancing procedure that operates individually for each classifier in the pool, leveraging insights from~\cite{Wood}\cite{DiversityMeasure}, which highlight the importance of diversity among classifiers for improving ensemble performance.

Our primary contributions include: (1) a balanced procedure for ensemble learning that enhances diversity by training each classifier in the pool with different random sampling and replacement of the training set; (2) evaluation of various machine learning algorithms from monolithic classifiers, static, and dynamic ensemble learning, using diverse metrics; and (3)~an instance hardness analysis, demonstrating reduced dataset difficulty after balancing.
%%%%%%%%%%%%%%%%%%%%%%%%%%%%%%%%%%%%%%%%%%%%%%%%%%%%%%%%%%%%%%%%%%%%%%%%%%%%%%%%%%%%%%%%%%%%%%%%%%%%%%%%%
\section{Related Work}  

%Malware detection is a critical challenge in information security, where identifying malicious samples can be complicated by the disproportion between classified malicious and benign data. This class imbalance can lead to misleading performance metrics, making it essential to apply appropriate techniques to balance the classes. Imbalanced data can lead to misleading performance metrics, highlighting the need for a comprehensive examination of balancing techniques~\cite{HuayanayImbalanced}.
The class imbalance in malware detection complicates the identification of malicious samples and can distort performance metrics. This challenge makes it crucial for a comprehensive examination of balancing techniques~\cite{HuayanayImbalanced}.

%The Drebin dataset \cite{Drebin} has been extensively used in mobile security research, particularly in the context of malware detection on Android systems, as highlighted in several studies \cite{fast}\cite{koreaScience}\cite{DosAndDonts}. These researchers have explored different approaches to classifying malicious and benign applications using this dataset as a reference due to its wide range of features and the significant number of malware samples. However, none of the works we reviewed that involve the Drebin dataset explicitly address the impact of class imbalance. Most analyses opt to use only a subset of the dataset.

The Drebin dataset \cite{Drebin} is extensively used in Android malware detection research~\cite{fast}\cite{koreaScience}\cite{DosAndDonts} due to its wide range of features and the significant number of malware samples. However, studies using this dataset often overlook class imbalance issues, frequently analyzing only subsets of the data.

\noindent \textbf{Machine Learning-Based Approaches}.
Arp et al. \cite{Drebin}, creators of the Drebin dataset, proposed a method combining static analysis and Support Vector Machines (SVM) to detect Android malware, achieving a high detection rate. Later, Wang et al. \cite{wang} demonstrated that using deep neural networks with features from the Drebin dataset enhances detection accuracy compared to traditional methods.

% One of the pioneering studies utilizing the Drebin dataset was conducted by Arp et al. \cite{Drebin}, the creators of the dataset itself. In this work, the authors proposed a method that combines static analysis with machine learning to detect malware on Android. They used a classifier based on Support Vector Machines (SVM) and achieved a high detection rate, demonstrating the effectiveness of supervised learning techniques in malware identification.

% More recently, Wang et al. \cite{wang} explored the use of deep neural network for malware detection on Android devices. Their study demonstrated that combining deep learning techniques with features extracted from the Drebin dataset can improve detection accuracy compared to traditional approaches.

\noindent \textbf{Ensemble Techniques and Hybrid Methods.}
Ensemble and hybrid methods have been explored to enhance classifier performance. Suarez-Tangil et al. \cite{fast} introduced a detection system combining bagging and boosting with monolithic classifiers like Decision Trees and SVM, demonstrating on the Drebin dataset that ensemble techniques improve robustness and reduce performance variability across different experimental conditions. Similarly, Zhou et al. \cite{zhou} showed that hybrid methods integrating static and dynamic analysis achieve higher detection rates, particularly on large datasets like Drebin.

% Other studies have focused on applying ensemble methods and hybrid approaches to improve classifier performance. For example, Suarez-Tangil et al. \cite{fast} proposed a detection system that combines bagging and boosting techniques, applying these methods to monolithic classifiers such as Decision Trees and SVM. Using the Drebin dataset, the authors demonstrated that ensemble techniques can increase the model's robustness by maintaining low variability in performance, even across different experimental runs with varying random seeds.

% Additionally, Zhou et al. \cite{zhou} investigated the application of hybrid methods that combine static and dynamic analysis for malware detection. They showed that combining different analysis techniques can result in higher detection rates, especially when applied to large datasets like Drebin.

\noindent \textbf{Challenges and Limitations.}  
The main challenges in malware detection include limited availability of large, diverse datasets, and imbalance between benign and malicious samples. %, and ineffective dynamic selection techniques. 
Small datasets hinder model generalization across malware types \cite{ImportantSize}, limiting real-world effectiveness. Imbalance biases classifiers toward the majority class, making proper data balancing essential for fair evaluation \cite{Bifet}. Additionally, using robust evaluation metrics is crucial for accurately assessing model performance, especially in difficult-to-classify cases. This work addresses these challenges to improve malware detection model robustness.
%The main challenges in malware detection stem from the limited availability of large and diverse datasets, the imbalance between benign and malicious samples, and the lack of effective dynamic selection techniques. Small dataset sizes hinder a model's ability to generalize across different malware types\cite{ImportantSize}, thus limiting its real-world effectiveness. Dataset imbalance can make classifiers biased toward the majority class, making it crucial to properly balance and evaluate the data for fair performance assessment\cite{Bifet}. Moreover, using robust evaluation metrics is key to accurately gauging model performance, particularly in difficult-to-classify cases, where poor handling can greatly diminish model reliability. In this work, we address these challenges to enhance the robustness of malware detection models.

\noindent \textbf{Contributions of the Present Study.}  
This study builds on the Drebin dataset, focusing on underexplored data balancing techniques and dynamic model selection. By comparing various machine learning methods, it provides a deeper analysis of how data balancing impacts malware detection models, offering valuable insights and contributions to the field.

%This study expands existing knowledge using the Drebin dataset, but focusing on data balancing techniques and dynamic model selection, which have been underexplored in the literature. By comparing the performance of various machine learning methods, this research aims to provide a deeper analysis of how data balancing affects malware detection models. The proposed framework not only enhances our understanding of these techniques but also offers a valuable contribution to the field.
%%%%%%%%%%%%%%%%%%%%%%%%%%%%%%%%%%%%%%%%%%%%%%%%%%%%%%%%%%%%%%%%%%%%%%%%%%%%%%%%%%%%%%%%%%%%%%%%%%%%%%%%%

\section{The proposed framework}

\begin{figure}
    \centering
    \includegraphics[scale=0.57]{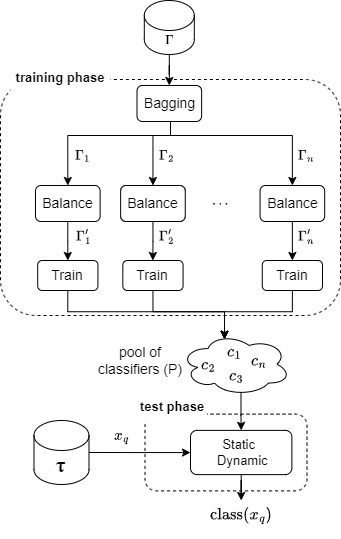}
    \caption{\textbf{Experimental framework for empirical evaluation of malware detection models on android devices.}}
    \label{fig:rotulo}
\end{figure}

In our approach, Bootstrap-Based Balancing, we apply a balancing technique individually to each bootstrap sample. This aims to increase variability across the classifiers, ensuring that each classifier is exposed to a different distribution of the data. As a result, the classifiers in the pool become more diverse, which, in turn, strengthens the overall robustness of the ensemble \cite{Wood}\cite{DiversityMeasure}. Figure~\ref{fig:rotulo} shows the proposed framework, composed of two phases: training and testing. The training phase generates a pool of classifiers ($P$) given a training dataset ($\Gamma$). The testing phase aims to classify a query instance ($x_q$) using $P$. %These two phases are detailed in the following sections.

\noindent \textbf{Training Phase.}
Given the training dataset ($\Gamma$), this phase generates a diverse pool of classifiers by first applying ``Bagging'', a random procedure that creates $n$ bootstraps (${\Gamma_1, \Gamma_2, \ldots, \Gamma_n}$) with replacement. Widely used in ensemble learning, Bagging consistently leads to satisfactory results~\cite{pastor15,DESRecentAdvances}.

%Given the training dataset ($\Gamma$), this phase aims to generate a diverse pool of classifiers. Thus, the first step is to apply ``Bagging'', which is a random procedure that generate $n$ bootstraps ($\{\Gamma_1, \Gamma_2, \ldots, \Gamma_n\}$) with replacement. Bagging is commonly used in ensemble learning, leading to satisfactory results~\cite{pastor15,DESRecentAdvances}. 

After, each bootstrap ($\Gamma_i$) is balanced separately, similarly as performed by Roy et al.~\cite{Anandarup}, generating $n$ bootstraps ($\{\Gamma_1', \Gamma_2', \ldots, \Gamma_n'\}$). So, instead of balancing the whole training dataset ($\Gamma$), this procedure aims to increase the level of diversity since each $\Gamma_i$ is processed independently. The random sampling and class balancing introduce noise, leading to distinct data distributions across the subsets.

The last step is to train $n$ classifiers, each one using a different dataset ($\Gamma_i$), composing the pool of classifiers ($P = \{c_1, c_2, \ldots, c_n\}$), i.e., $c_i = train(\Gamma_i), \quad \forall i \in \{1, 2, \ldots, n\}$. Since each classifier only has access to part of the whole training data, each classifier is expected to be an expert in a different region of the feature space.

\noindent \textbf{Testing Phase}
After the training phase, a pool of classifiers ($P = \{c_1, c_2, \ldots, c_n\}$) is used to predict the class of a query instance $x_q \in \tau$. 
%The fusion of the classifiers answers can be performed by different classification approaches such as, static and dynamic combination of classifiers~\cite{DESRecentAdvances}.
Different classification approaches, such as static and dynamic combinations of classifiers can perform the fusion of the classifier's answers~\cite{DESRecentAdvances}.

To the best of our knowledge, dynamic selection (DS) algorithms have not been previously evaluated for imbalanced malware classification. The variability across different regions of the feature space makes it difficult for static methods to perform consistently well in all cases. DS algorithms are particularly appealing in this context because they adaptively select the most relevant classifiers for each instance, addressing the challenge of imbalanced data. Specifically, DS algorithms select a subset of the most competent classifiers from the pool ($P$) for each query instance ($x_q$), doing so on-the-fly, i.e., during the generalization phase. This approach assumes that different classifiers excel in different local regions of the feature space, allowing for more specialized predictions.

%%%%%%%%%%%%%%%%%%%%%%%%%%%%%%%%%%%%%%%%%%%%%%%%%%%%%%%%%%%%%%%%%%%%%%%%%%%%%%%%%%%%%%%%%%%%%%%%%%%%%%%%%
\section{Methodology of the Experiments}

%All models were evaluated across 30 iterations to account for statistical variability. For each iteration, the dataset was randomly split into 80\% for training and 20\% for testing. This process was repeated to mitigate the effects of random partitioning on model performance.

%This section presents information related to the experimental study, such as: dataset (Section~\ref{sc:dataset}), learning algorithms (Section~\ref{sc:LearningAlgorithms}) and balancing algorithms (Section~\ref{sc:smote}), as well as the metrics (Section~\ref{sc:Metrics}) and the experimental environment (Section~\ref{sc:environment}).

\noindent   \textbf{Dataset.}\label{sc:dataset}
The DREBIN database \cite{Drebin}, designed for malware detection, contains 129,013 instances, of which 5,560 are malware. The imbalance ratio (IR) is 22.20, presenting a significant challenge due to its highly imbalanced nature\cite{IRDef}.

DREBIN extracts features from applications, capturing behavioral and structural aspects across eight categories: \textbf{hardware components} used or interacted by the application, \textbf{requested permissions}, \textbf{application components} (e.g., activities, services), \textbf{filtered intents} handled by the application, \textbf{restricted API calls}, \textbf{permissions} actively used during execution, \textbf{suspicious API calls} associated with known malware behavior, and \textbf{network addresses}. The data and feature sets are available through \cite{LinkDrebin}.

All models were evaluated across 30 iterations, with each iteration using a random split of 80\% for training and 20\% for testing to account for statistical variability.

% Each of the eight defined categories corresponds to a specific aspect of the code being analyzed. When a particular category is identified during the analysis, the count of features within that group is incremented accordingly. For example, if an application has access to 2 hardware components, requests 11 permissions, contains 5 distinct components, filters 3 intents, calls 7 restricted APIs, uses 6 permissions, invokes 11 suspicious APIs, and communicates with 26 network addresses, the resulting feature vector would be (2, 11, 5, 3, 7, 6, 11, 26). The data, application samples, and all extracted feature sets can be accessed through the link provided by the authors of the study \cite{LinkDrebin}.

% \noindent \textbf{Data Balancing.}\label{sc:smote}
% Given the class imbalance in the DREBIN dataset, traditional machine learning models often favor predictions for the majority class, leading to poor generalization and inflated accuracy metrics \cite{ImbalancedReview} \cite{Bifet}. To address this challenge, we utilized the SMOTE (\textit{Synthetic Minority Over-sampling Technique}) for data balancing~\cite{SMOTE}, which generates synthetic examples for the minority class based on existing data. This approach helps prevent bias towards the majority class and reduces the risk of overfitting associated with simply duplicating minority instances. The SMOTE hyperparameters ensure that the synthetic samples accurately reflect the minority class distribution, with the generation process considering five neighbors to maintain diversity among the newly created instances.

\noindent \textbf{Data Balancing.}\label{sc:smote}
The class imbalance in the DREBIN dataset causes traditional machine learning models to favor the majority class, resulting in poor generalization and inflated accuracy metrics \cite{ImbalancedReview} \cite{Bifet}. To mitigate this, we applied SMOTE (\textit{Synthetic Minority Over-sampling Technique})~\cite{SMOTE}. This method prevents majority class bias and reduces overfitting risks associated with duplicating instances. 

\noindent \textbf{Models.}\label{sc
} 
This study addresses binary malware detection using three classifier categories: monolithic, static ensemble, and dynamic selection. Monolithic classifiers (e.g., Decision Tree, KNN, MLP, Naive Bayes) were used directly and with Bagging (e.g., Bagging Decision Tree, KNN, MLP, Naive Bayes) to enhance robustness and retain variance information by aggregating predictions from bootstrap samples \cite{MonoliticSurvey,StaticEnsembleSurvey}, all implemented with scikit-learn 1.4.2 \cite{SklearnPaper}. We also propose ``Bootstrap-Based Balancing" (Figure~\ref{fig:rotulo}) (BBB) to preprocess bootstrap samples by generating synthetic minority instances for better class balance and diversity \cite{Anandarup}.

Static ensemble classifiers (e.g., Gradient Boosted Decision Tree, Random Forest, Single Best, Static Selection) use a fixed ensemble \cite{RFandGBSurvey,DESRecentAdvances}. Dynamic selection classifiers (e.g., KNOP, METADES, OLA) adaptively select the best models for each instance. Implemented with DESlib 0.3.7 \cite{DESlibPaper}. The classifiers hyperparameters are detailed on GitHub.

\noindent \textbf{Evaluation Metrics.}\label{sc:Metrics}
Since we are dealing with highly imbalanced data \cite{IRDef}, we must choose metrics that provide a comprehensive view of the model's performance across both classes. Metrics such as accuracy can be misleading, as they may give a false sense of high performance when, in reality, the minority class is being misclassified \cite{Metricas de Avaliação}. Therefore, we will use recall, F1 score, G-Mean and Matthews Correlation Coefficient (MCC) for evaluation~\cite{HuayanayImbalanced}, as these metrics are robust to imbalanced datasets.

%%%%%%%%%%%%%%%%%%%%%%%%%%%%%%%%%%%%%%%%%%%%%%%%%%%%%%%%%%%%%%%%%%%%%%%%%%%%%%%%%%%%%%%%%%%%%%%%%%%%%%%%%
\section{Results and Discussion}

This section presents the experimental results for both the original and balanced training sets, alongside an assessment of data hardness and its impact on classifier performance. Table~\ref{tab:Combined_metrics} summarizes the performance metrics for the classifiers before and after balancing, with color coding highlighting the top performances: red (best), blue (second), and burgundy (third).

\noindent \textbf{Experiments with Original Dataset.}  
Among the classifiers, Decision Tree excelled in Recall and G-Mean, effectively identifying positive instances due to its hierarchical structure \cite{MonoliticSurvey}. In contrast, Naive Bayes and MLP underperformed, with Naive Bayes showing the lowest F1 Score, highlighting challenges for single classifiers with imbalanced datasets. Bagging Decision Tree improved F1 Score over the standalone version, but gains were limited, indicating ensemble benefits depend on the base model's strength.

Static ensemble models, like Random Forest and Static Selection, achieved high performance, demonstrating the effectiveness of ensemble strategies, while GBDT underperformed due to its iterative approach, which can bias against minority classes \cite{RFandGBSurvey}. Dynamic selection models consistently excelled in F1 Score and MCC, leveraging adaptive selection to handle imbalanced data. Overall, ensemble and dynamic methods outperformed monolithic classifiers, with Decision Tree and dynamic selectors delivering robust, consistent results.

\noindent \textbf{Experiments after Balancing.} 
The study compared monolithic, static, bagging, and dynamic ensemble algorithms for balancing and classifying datasets, with monolithic models using traditional balancing and bagging/dynamic models employing the novel BBB technique. Balancing improved minority class recall and G-Mean but slightly decreased precision metrics like F1 score and MCC, likely due to false positives. Among monolithic models, KNN excelled in recall and G-Mean, while Naive Bayes underperformed. MLP showed competitive results but faced precision-recall trade-offs.

Bagging outperformed monolithic models, improving F1 score and MCC despite small declines in recall and G-Mean. GBDT followed similar trends with gains in recall and G-Mean but lower F1 score and MCC. Static selectors performed similarly after balancing, with minor trade-offs. Dynamic selectors, particularly KNOP, showed the best overall performance, adapting to data characteristics and excelling across all metrics. In conclusion, dynamic selection with BBB outperformed monolithic, bagging, and static selectors, offering greater robustness and reduced bias.

\begin{table*}[h]
    \centering
    \setlength{\tabcolsep}{3pt}  
    \renewcommand{\arraystretch}{1.1}
\caption{\textbf{Performance metrics of various classifiers before and after balancing, with standard deviations shown in parentheses.}}
    \label{tab:Combined_metrics}
    \begin{tabular}{l|llll|llll}
    \toprule

\textbf{Model} & \multicolumn{4}{c|}{\textbf{Imbalanced}} & \multicolumn{4}{c}{\textbf{Balanced}} \\
               & \textbf{Recall} & \textbf{F1 score} & \textbf{G-Mean} & \textbf{MCC} & \textbf{Recall} & \textbf{F1 score} & \textbf{G-Mean} & \textbf{MCC} \\
    \midrule
DecisionTree    &  \textcolor{red}{82.24(1.19)}             & 79.65(0.94)    &  \textcolor{red}{90.19(0.65)}    & 78.75(0.98)   & 86.11(0.82)                          & 71.02(0.32) & 91.61(0.40)                      & 70.69(0.29)  \\
KNN             & 73.42(1.70)                              & 77.11(1.20) & 85.35(0.99)                      & 76.25(1.23)   & \textcolor{red}{89.30(0.82)}      & 63.93(0.54) & 92.56(0.38)                      & 64.82(0.47) \\
MLP             & 59.26(4.40)                              & 68.51(2.38) & 76.69(2.85)                      & 68.37(1.89)   & 88.55(1.63)                          & 53.28(3.12)           & 90.96(0.48)                      & 55.53(2.43) \\
NaiveBayes      & 46.27(1.77)                              & 32.52(1.18) & 65.86(1.23)                      & 30.06(1.27)   & 73.72(3.73)                          & 18.85(0.27) & 73.11(1.13)                      & 20.65(0.71) \\
\hline
BaggingDT       &  \textcolor{purple}{81.15(1.08)}         & 85.07(0.67) &  \textcolor{blue}{89.89(0.59)}   & 84.54(0.69)   & 86.00(0.35) & \textcolor{purple}{81.64(0.52)} & 92.22(0.19) & \textcolor{purple}{80.88(0.52)} \\
BaggingKNN      & 73.19(1.31)                              & 77.36(1.04) & 85.24(0.76)                      & 76.54(1.08)   & \textcolor{blue}{88.99(0.46)}                      & 69.69(0.31) & \textcolor{red}{92.91(0.23)} & 69.87(0.32) \\
BaggingMLP      & 60.76(1.71)                              & 71.17(1.17) & 77.76(1.09)                      & 71.26(1.07)   & \textcolor{purple}{88.80(1.09)} & 59.93(1.00) &91.92(0.56) &61.26(0.97)\\
BaggingNB       & 46.53(1.23)                              & 32.83(0.81) & 66.07(0.85)                      & 30.38(0.87)   & 73.97(0.21) & 18.70(0.20) & 73.08(0.15) & 20.51(0.21) \\
\hline
GBDT            & 46.93(2.04)                              & 60.29(1.83) & 68.36(1.49)                      & 61.78(1.63)   & 86.65(1.37)                      & 42.04(0.73) & 88.23(0.69) & 45.52(0.83) \\
RandomForest    & 80.45(1.28)                              & 86.43(0.84) & 89.58(0.71)                      & 86.12(0.84)  & 87.40(0.89)                      & 79.13(0.84) & \textcolor{blue}{92.78(0.47)} & 78.49(0.85) \\
SingleBest      & 79.00(1.03)                              & 77.71(0.79) & 88.39(0.57)                      & 76.71(0.83)                   & 86.89(0.96)                      & 67.01(0.88) & 91.68(0.50) & 67.16(0.86) \\
StaticSelection & 80.39(1.12)                              &  \textcolor{purple}{86.52(0.78)} & 89.55(0.62)                      &  \textcolor{purple}{86.24(0.79)}  & 87.19(0.91)                      & 79.41(0.81) & \textcolor{purple}{92.69(0.47)} & 78.74(0.82) \\
\hline

KNOP            & 80.49(1.11)                              &  \textcolor{red}{86.69(0.68)}     & 89.61(0.62)                      &  \textcolor{red}{86.43(0.68)} & 84.32(1.07) & \textcolor{red}{84.58(0.87)} & 91.51(0.55) & \textcolor{red}{83.91(0.92)} \\
METADES         &  80.92(1.06)                             &  \textcolor{blue}{86.60(0.63)} & 89.83(0.58)                      &  \textcolor{blue}{86.27(0.64)}    & 83.47(1.27) & \textcolor{blue}{83.40(0.84)} & 91.01(0.67) & \textcolor{blue}{82.67(0.88)} \\
OLA             & \textcolor{blue}{81.57(0.96)} & 90.29(0.47)                      & \textcolor{purple}{89.87(0.52)}                      & 79.21(0.73)   & 84.85(1.00) & 74.59(0.75) & 91.22(0.52) & 73.91(0.76) \\
\bottomrule
\end{tabular}
\end{table*}

\begin{figure}
    \centering
    \includegraphics[scale=0.55]{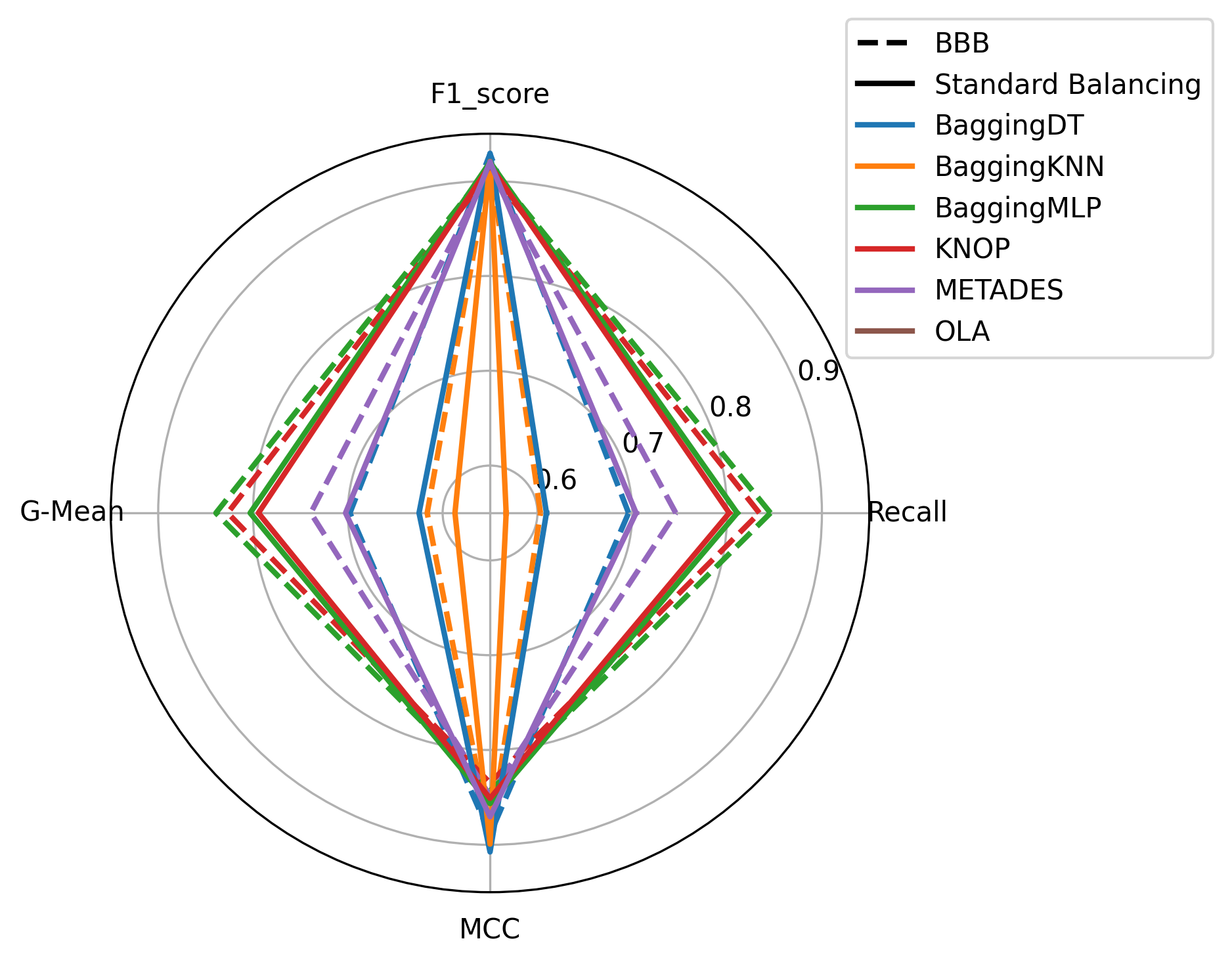}
    \caption{\textbf{Performance comparison of balancing techniques and models.}}
    \label{fig:balancing}
\end{figure}

\noindent \textbf{Comparison of Balancing Methods.}  
Figure~\ref{fig:balancing} compares static ensemble (top) and dynamic selection (bottom) methods under two balancing techniques: BBB and standard balancing. The proposed method outperforms traditional balancing, particularly in G-Mean and recall, indicating that applying SMOTE more granularly to each training subset reduces false negatives in both static and dynamic models. While BBB shows a greater improvement in static ensembles, its impact on dynamic selection is less pronounced but still noticeable.

\begin{figure}[h]
    \centering
    \includegraphics[scale=0.24]{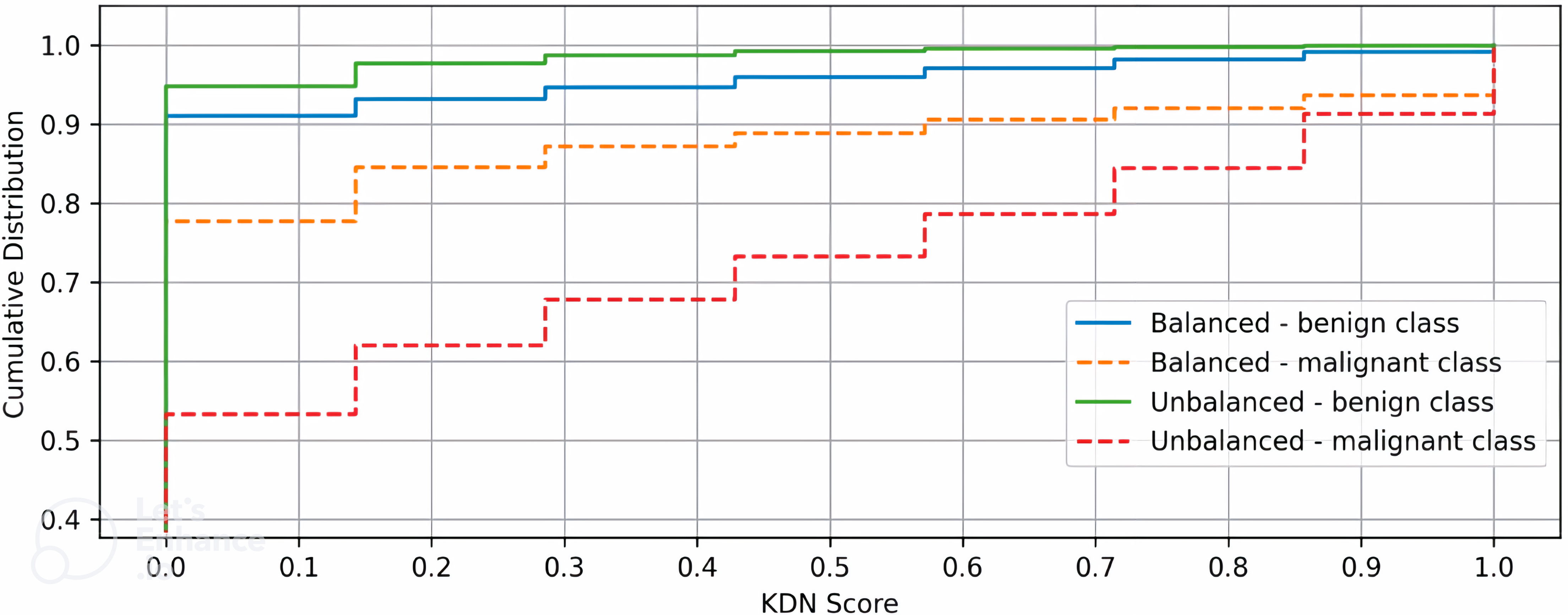} 
    \caption{\textbf{Cumulative distribution of KDN score.}}
    \label{fig:IH}
\end{figure}

\noindent \textbf{Exploring Instance Hardness.} To better understand the results, instance hardness was assessed using the KDN (K-Disagreeing-Neighbors) method, as shown in Figure~\ref{fig:IH}. In the imbalanced scenario, benign instances are easier to classify, with many having low KDN scores, while malignant instances have higher KDN scores, indicating greater difficulty. This aligns with the average hardness values: the benign class has an average hardness of 0.0148, while the malignant class has a significantly higher average of 0.2704.

Balancing affects KDN scores by slightly increasing the classification difficulty for benign instances (average hardness rises to 0.0440) and making malignant instances easier to classify (average hardness drops to 0.1221). This occurs because balancing reduces the isolation of malignant instances, improving their classification performance, while benign instances face more neighbors from the opposite class.

These changes impact performance metrics like recall and F1 score. Before balancing, benign instances had high recall, but malignant instances suffered due to higher hardness. After balancing, the recall of the malignant class improves significantly, though at the expense of reduced performance for the benign class, reflected in lower F1 score and MCC.

\section{Conclusion and Future Work}

In conclusion, our study demonstrates that enhancing the performance of the minority class often involves trade-offs with the majority class, as balancing techniques, while improving minority recall, can reduce precision, as seen in declines in F1 score and MCC. Dynamic selection methods and the proposed Bootstrap-Based Balancing technique proved effective, consistently delivering robust performance by enabling classifiers to adapt to the complexity of this highly imbalanced dataset. Our analysis of instance hardness further highlights the relationship between class balance and classification difficulty, with shifts in KDN scores post-balancing showing how strategic adjustments can benefit both classes. For future work, we aim to enhance minority class performance while minimizing trade-offs for the majority class through more robust balancing techniques.
%%%%%%%%%%%%%%%%%%%%%%%%%%%%%%%%%%%%%%%%%%%%%%%%%%%%%%%%%%%%%%%%%%%%%%%%%%%%%%%%%%%%%%%%%%%%%%%%%%%%%%%%%


\begin{thebibliography}{00}

\bibitem{Drebin} Daniel Arp et al. "Drebin: Efficient and Explainable Detection of Android Malware in Your Pocket", Annual Network and Distributed System Security Symposium (NDSS), 2014.


\bibitem{LinkDrebin} https://drebin.mlsec.org/

\bibitem{koreaScience} Luo Shi-qi et al. "Deep Learning in Drebin: Android malware Image Texture Median Filter Analysis and Detection,” KSII Transactions on Internet and Information Systems, vol. 13, no. 7, 2019.


\bibitem{Metricas de Avaliação} De Diego et al. General Performance Score for classification problems. Applied Intelligence 52, 12049–12063 (2022)

\bibitem{SMOTE} Chawla et al. (2002). SMOTE: Synthetic Minority Over-sampling Technique. Journal of Artificial Intelligence Research, 16, 321-357.

\bibitem{wang} Wang et al. (2019). DroidEnsemble: Detecting Android malicious applications with ensemble of string-based features. Journal of Computer Virology and Hacking Techniques, 15(1), 29-38.

\bibitem{fast} Suarez-Tangil et al. (2017). DroidSieve: Fast and Accurate Classification of Obfuscated Android Malware. In ACM Conference on Data and Application Security and Privacy, 309-320.

\bibitem{zhou}  Zhou et al. (2012). Dissecting Android malware: Characterization and evolution. IEEE Symposium on Security and Privacy.

\bibitem{MLSurveyDetection} Daniele Ucci et al, "Survey of machine learning techniques for malware analysis", Computers \& Security, Volume 81, Pages 123-147, 2019.

\bibitem{MalwareDefinition} Imtithal A. Saeed et al. "A Survey on Malware and Malware Detection Systems", International Journal of Computer Applications, Volume 67, Number 16, 2013.

\bibitem{MalwareAnalysisAproach} Sihwail, Ramiet al. A Survey on Malware Analysis Techniques: Static, Dynamic, Hybrid and Memory Analysis. International Journal on Advanced Science, Engineering and Information Technology, 2018.

\bibitem{GlobalRiskReport} World Economic Forum, The Global Risks Report 2024. Accessed: Sep. 18, 2024. [Online]. Available: https://www.weforum.org/publications/global-risks-report-2024/

\bibitem{ImbalancedReview} Yanminsun et al. (2011). Classification of imbalanced data: a review. International Journal of Pattern Recognition and Artificial Intelligence.

\bibitem{DosAndDonts}Arp, Daniel et al. Dos and Don’ts of Machine Learning in Computer Security. In: USENIX Security Symposium, 2022.

\bibitem{Anandarup}Anandarup Roy et al. A study on combining dynamic selection and data preprocessing for imbalance learning, Neurocomputing, Volume 286, 2018

\bibitem{HuayanayImbalanced}Huayanay, Alex de la Cruz et al. Performance of evaluation metrics for classification in imbalanced data. Computational Statistics, 2024.

\bibitem{MonoliticSurvey}Kotsiantis, S. B. et al. Supervised machine learning: A review of classification techniques. Informatica, 249-268, 2007.

\bibitem{StaticEnsembleSurvey}Rokach, Lior. Ensemble-based classifiers. Artificial Intelligence Review, Dordrecht, v. 33, n. 1, p. 1–39, 2010.

\bibitem{RFandGBSurvey}Dong, Xibin et al. A survey on ensemble learning. Frontiers of Computer Science, v. 14, n. 2, p. 241-258, 2020.

\bibitem{DESRecentAdvances}Cruz, R. M. O. et al. Dynamic classifier selection: Recent advances and perspectives. Information Fusion, v. 41, p. 195-216, 2018. 

\bibitem{pastor15}José F. Díez-Pastor et al. Diversity techniques improve the performance of the best imbalance learning ensembles, Information Sciences, v. 325, p. 98-117, 2015.

\bibitem{Aslan}Aslan, Ö. et al. A comprehensive review of cyber security vulnerabilities, threats, attacks, and solutions. Electronics, 12(6), 1333.

\bibitem{Bifet} Bifet, A. et al. ”A Survey on Learning from Imbalanced Data Streams”, Journal of Machine Learning Research, 2022.

\bibitem{Wood} Wood, D. et al. "A Unified Theory of Diversity in Ensemble Learning", Journal of Machine Learning Research, vol. 24, 2023.

\bibitem{Cruz}Cruz, R. M. O. et al. Dynamic classifier selection: Recent advances and perspectives. Information Fusion, 41, 195–216.

\bibitem{DiversityMeasure}Kuncheva, L. I. et al. Measures of diversity in classifier ensembles and their relationship with the ensemble accuracy. Machine Learning, 51(2), 181–207.

\bibitem{IRDef}Khan, A. A. et al. A review of ensemble learning and data augmentation models for class imbalanced problems: Combination, implementation and evaluation. Expert Systems With Applications, 244, 122778.

\bibitem{DESlibPaper} Rafael M. O. Cruz et al. DESlib: A Dynamic ensemble selection library in Python, Journal of Machine Learning Research, 2020.

\bibitem{SklearnPaper} Pedregosa et al. Scikit-learn: Machine Learning in Python, JMLR 12, pp. 2825-2830, 2011.

\bibitem{ImportantSize}Sun, C. et al. "Revisiting Unreasonable Effectiveness of Data in Deep Learning Era". International Conference on Computer Vision, 2017.

\end{thebibliography}
\end{document}